\documentclass[journal,twoside]{IEEEtran}

\usepackage{cite}

\ifCLASSINFOpdf
   \usepackage[pdftex]{graphicx}
   \DeclareGraphicsExtensions{.pdf,.jpeg,.png}
\else
\fi

\usepackage{amsmath}
\usepackage{array}

\usepackage{fixltx2e}

\usepackage{url}

\usepackage{times}
\usepackage{epsfig}
\usepackage{graphicx}
\usepackage{float}
\usepackage{wrapfig}
\usepackage{amsmath,amssymb,amsthm}
\usepackage{bm,xspace}
\usepackage{comment}
\usepackage{verbatim}
\usepackage{multirow}
\usepackage{balance}
\usepackage{url}
\usepackage{booktabs}
\usepackage{etoolbox,siunitx}
\usepackage{calc}
\usepackage{pifont,hologo}
\usepackage{nicefrac}

\newcommand{\ra}[1]{\renewcommand{\arraystretch}{#1}}

\usepackage[super]{nth}
\usepackage{nicefrac}
\robustify\bfseries

\def\nn{\mathbf{n}}

\renewcommand{\ss}{\mathbf{s}}
\def\tt{\mathbf{t}}

\def\xx{\mathbf{x}}

\def\KK{\mathbf{K}}
\def\LL{\mathbf{L}}

\def\lL{\mathcal{L}}

\DeclareMathSymbol{@}{\mathord}{letters}{"3B}

\newcommand\timess{\mathbin{\!\times\!}}

\newcommand\mypara[1]{\paragraph{#1}}

\def\latex/{\LaTeX}
\def\bibtex/{\hologo{BibTeX}}

\begin{document}

\title{Speech Denoising with Deep Feature Losses}

\author{Fran\c{c}ois~G.~Germain,~%
        Qifeng~Chen,~%
        and~Vladlen~Koltun%
\thanks{F. Germain is with the Center for CCRMA, Stanford University, Stanford, CA 94305. E-mail: \texttt{francois@ccrma.stanford.edu}. This work was performed while he was interning at Intel Labs. Q. Chen and V. Koltun are with the Intelligent Systems Lab, Intel Labs, Santa Clara, CA 95054.}}
\markboth{Preprint}%
{Preprint: Germain \MakeLowercase{\textit{et al.}}: Speech Denoising with Deep Feature Losses}

\maketitle

\begin{abstract}
	We present an end-to-end deep learning approach to denoising speech signals by processing the raw waveform directly. Given input audio containing speech corrupted by an additive background signal, the system aims to produce a processed signal that contains only the speech content. Recent approaches have shown promising results using various deep network architectures. In this paper, we propose to train a fully-convolutional context aggregation network using a deep feature loss. That loss is based on comparing the internal feature activations in a different network, trained for acoustic environment detection and domestic audio tagging. Our approach outperforms the state-of-the-art in objective speech quality metrics and in large-scale perceptual experiments with human listeners. It also outperforms an identical network trained using traditional regression losses. The advantage of the new approach is particularly pronounced for the hardest data with the most intrusive background noise, for which denoising is most needed and most challenging.
\end{abstract}


\IEEEpeerreviewmaketitle

\section{Introduction} \label{sec:introduction}

\IEEEPARstart{S}{peech} denoising (or enhancement) refers to the removal of background content from speech signals \cite{Loizou2013}. Due to the ubiquity of this audio degradation, denoising has a key role in improving human-to-human (e.g., hearing aids) and human-to-machine (e.g., automatic speech recognition) communications. A particularly challenging but common form of the problem is the under-determined case of single-channel speech denoising, due to the complexity of speech processes and the unknown nature of the non-speech material. The complexity is further compounded by the nature of the data, since audio material contains a high density of data samples (e.g., 16,000 samples per second). Challenges also arise in mediated human-to-human communication, as perception mechanisms can make small errors still noticeable by the average user~\cite{BosiGoldberg2002}.

In this work, we present an end-to-end deep learning approach to speech denoising. Our approach trains a fully-convolutional denoising network using a deep feature loss. To compute the loss between two waveforms, we apply a pretrained audio classification network to each waveform and compare the internal activation patterns induced in the network by the two signals. This compares a multitude of features at different scales in the two waveforms. We perform extensive experiments that compare the presented approach to recent state-of-the-art end-to-end deep learning techniques for denoising. Our approach outperforms them in both objective speech quality metrics and large-scale perceptual experiments with human listeners, which indicate that our approach is more effective than the baselines. The advantages of the presented approach are particularly pronounced for the hardest, noisiest inputs, for which denoising is most challenging.

\subsection{Related Work}
\label{sec:related}

Before the popularization of deep networks, denoising systems relied on spectrogram-domain statistical signal processing methods~\cite{Loizou2013}, followed more recently by spectrogram factorization-based methods~\cite{Smaragdis2014}. Current denoising pipelines instead rely on deep networks for state-of-the-art performance. However, most pipelines still operate in the spectrogram domain~\cite{Wang2012,Lu2013,Narayanan2013,Weninger2014,Xu2015,Kumar2016,ZhangWang2016,ChenWang2017}. As such, signal artifacts then arise due to time aliasing when using the inverse short-time Fourier transform to produce the time-domain enhanced signal. This particular issue can be somewhat alleviated, but with increased computational cost and system complexity~\cite{RouxVincent2013,Germain2014,Gerkmann2015,Wang2015,Erdogan2015,Williamson2017,Moorer2017}.

Recently, there has been growing interest in the design of performant denoising pipelines that are optimized end-to-end and directly operate on the raw waveform. Such approaches aim at fully leveraging the expressive power of deep networks while avoiding expensive time-frequency transformations or loss of phase information~\cite{Fu2017,Rethage2017,Pascual2017,Qian2017}. Some of these approaches typically use simple regression loss functions for training the network~\cite{Fu2017,Rethage2017} (e.g., $L^1$ loss on the raw waveform), while ones with more advanced loss functions have shown limited gains in mismatched conditions~\cite{Pascual2017,Qian2017}.

For our loss function, we are inspired by computer vision research, where activations in pretrained classification networks were found to yield effective loss functions for image stylization and synthesis~\cite{Johnson2016,ChenKoltun2017}. To compute the loss between two images, these approaches apply a pretrained image classification network to both. Each image induces a pattern of internal activations in the network to be compared, and the loss is defined in terms of their dissimilarity. Such complex training losses have been shown to yield state-of-the-art algorithms without the need for prior expert knowledge or added complexity for the processing network itself. Furthermore, increased performance can be achieved even without task-specific loss networks~\cite{Zhang2018}. Our work develops this idea in the context of speech processing.

\section{Method}
\label{sec:overview}

\subsection{Denoising Network}
\label{ssec:denoising}

Let $\xx$ be an audio signal corresponding to speech $\ss$ that is corrupted by an additive background signal $\nn$ so that
$\xx = \ss + \nn$. Our goal is to find a denoising operator $g$ such that $g(\xx) \approx \ss$. We use a fully-convolutional network architecture based on context aggregation networks~\cite{YuKoltun2016}. The output signal is synthesized sample by sample as we slide the network along the input. Context aggregation networks have been previously used in the WaveNet architecture for speech synthesis \cite{Oord2016}. Our architecture is simpler than WaveNet~-- no skip connections across layers, no conditioning, no gated activations~-- while our loss function is more advanced, as described in Section~\ref{ssec:loss}.

\mypara{Context aggregation} Our network consists of $16$ convolutional layers. The first and last layers (the degraded input signal and the enhanced output signal, respectively) are \mbox{1-dimensional} tensors of dimensionality $N\timess 1$. The number of samples $N$ in the input signal varies and is not given in advance. The signal sampling frequency $f_s$ is assumed to be \SI{16}{\kilo\hertz}. Each intermediate layer is a \mbox{2-dimensional} tensor of dimensionality $N \timess W$, where $W$ is the number of feature maps in each layer. (We set $W=64$.) The content of each intermediate layer is computed from the previous layer via a dilated convolution with $3 \timess 1$ convolutional kernels~\cite{YuKoltun2016} followed by an adaptive normalization (see below) and a pointwise nonlinear leaky rectified linear unit (LReLU)~\cite{Maas2013} $\max(0.2 x, x)$. Because of the normalization, no bias term is used for the intermediate layers. We zero-pad all layers so that their ``effective'' length is constant at $N$. Our network is then trained to handle the beginning and end of audio files even when speech content is near the sequence edges.

The dilation operator aggregates long-range contextual information without changing sampling frequency across layers~\cite{YuKoltun2016,Oord2016}. Here, we increase the dilation factor exponentially with depth from $2^{0}$ for the 1st intermediate layer to $2^{12}$ for the 13th one. We do not use dilation for the 14th and last one. For the output layer, we use a linear transformation ($1\timess 1$ convolution plus bias with no normalization and no nonlinearity) to synthesize the sample of the output signal. The receptive field of the pipeline is $2^{14}+1$ samples, i.e., about \SI{1}{\second} of audio for $f_s=\SI{16}{\kilo\hertz}$. We thus expect the system to capture context on the time scales of spoken words. A similar network architecture was shown to be advantageous in terms of compactness and runtime for image processing~\cite{Chen2017}.

\mypara{Adaptive normalization} The adaptive normalization operator used in our network matches the one proposed in~\cite{Chen2017} and improves performance and training speed. It adaptively combines batch normalization and identity mapping of the input $x$ as the weighted sum $\alpha_k x + \beta_k BN(x)$ (where $\alpha_k, \beta_k \in \mathbb{R}$ are scalar weights for the \mbox{$k$-th} layer and $BN$ is the batch normalization operator \cite{IoffeSzegedy2015}). The weights $\alpha,\beta$ are learned by backpropagation as network parameters.

\subsection{Feature loss}
\label{ssec:loss}

In our experiments, simple training losses (e.g., $L^1$) led to noticeably degraded output quality at lower signal-to-noise ratios (SNRs). The network seemed to improperly process low-energy speech information of perceptual importance. Instead, we train the denoising network using a deep feature loss that penalizes differences in the internal activations of a pretrained deep network that is applied to the signals being compared. By the nature of layered networks, feature activations at different depths in the loss network correspond to different time scales in the signal. Penalizing differences in these activations thus compares many features at different audio scales.

In computer vision, there are standard classification networks such as \mbox{VGG-19}~\cite{SimonyanZisserman2015}, pretrained on standard classification datasets such as ImageNet~\cite{Russakovsky2015}. Such standard classification networks do not exist in the audio processing field yet, so we design and train our own feature loss network.

\mypara{Feature loss network}
We design a simple audio classification network inspired by the VGG architecture in computer vision~\cite{SimonyanZisserman2015}, since it is known as a particularly effective feature loss architecture~\cite{Zhang2018}. The network consists of 15 convolutional layers with $3\timess 1$ kernels, batch normalization, LReLU units, and zero padding. Each layer is decimated by 2, halving the length of the subsequent layer compared to the preceding one. The number of channels is doubled every 5 layers, with 32 channels in the first intermediate layer. Each channel in the last feature layer is average-pooled to yield the output feature vector. The receptive field is $2^{15}-1$ samples. We train the network using backpropagation by feeding its output vector as features to one or more logistic classifiers with a cross-entropy loss for one or more classification tasks.

\mypara{Denoising loss function}
Let $\Phi^m$ be the \mbox{$m$-th} feature layer of the feature loss network, with layers at different depths corresponding to features with various time resolutions. The feature loss function is defined as a weighted $L^1$ loss on the difference between the feature activations induced in different layers of the network by the clean reference signal~$\ss$ and the output $g(\xx)$ of the denoising network being trained:
\begin{equation}
\lL_{\ss,\xx}(\theta) = \sum_{m=1}^{M} \lambda_m \left\| \Phi^m(\ss) - \Phi^m(g(\xx;\theta)) \right\|_{1},
\label{eq:perceptualloss}
\end{equation}
where $\theta$ are the parameters of the denoising network.
The weights $\lambda_m$ are set to balance the contribution of each layer to the loss. They are set to the inverse of the relative values of $\left\| \Phi^m(\ss) - \Phi^m(g(\xx;\theta)) \right\|_{1}$ after 10 training epochs. (For these first 10 epochs, the weights are set to 1.)

\section{Training}
\label{sec:training}

\subsection{Feature Loss}
\label{ssec:perceptuallosstrain}

\mypara{Tasks}
To generate a general-purpose feature loss network, we train it jointly on multiple audio classification tasks (only the logistic classifier parameters are trained as task-dependent). We use two tasks from the DCASE 2016 challenge~\cite{Mesaros2018}: the acoustic scene classification task and the domestic audio tagging task. In the first task, we are provided with audio files featuring various scenes (e.g., beach); the goal is to determine the scene type for each file. In the second task, we are given audio files featuring events of interest (e.g., child speaking); the goal is to determine which events took place in each file (with possibly multiple events in one file).

\mypara{Data} For the scene classification task, the training set \cite{Mesaros2016} consists of \mbox{30-second-long} audio files sampled at 44.1kHz, split among 15 different scenes (i.e., classes). As we need to develop a feature loss for the reduced sampling frequency of 16kHz, we resample the data. The audio files are stereo, so we split them into two mono files. The training set contains $2@340$ files. For the tagging task, the training set {\it CHiME-Home-refine} \cite{Foster2015} consists of \mbox{4-second-long} mono audio files sampled at 16kHz, with 7 different tags (i.e., labels). The training set contains $1@946$ files.

\mypara{Training}
Network weights are initialized with Xavier initialization~\cite{GlorotBengio2010}. We use the Adam optimizer~\cite{KingmaBa2015} with a learning rate of $10^{-4}$. The model is trained for $2@500$ epochs. In each epoch, we iterate over the training data for each task, alternating between files from each task. The order of the files is randomized independently for each epoch. The dataset for the first task is larger than the one for the second task, so we present some of the files in the second dataset (chosen at random) a second time to preserve strict alternation between tasks. 1 epoch consists of $4@680$ iterations (1 file per iteration). As a data augmentation procedure, we do not present entire clips,  but present a continuous section of minimal duration $2^{15}$ samples that is culled at random for each iteration.

\subsection{Speech Denoising}
\label{ssec:setrain}

\mypara{Data}
We use the noisy dataset made available in~\cite{ValentiniBotinhao2016}. To our knowledge, this is the largest available dataset for denoising that provides pre-mixed data with a clearly documented mixing procedure. It also has the benefit of being the dataset used in two recent works that we use as baselines. All details concerning the data can be found in~\cite{ValentiniBotinhao2016}. The training set is generated from the speech data of 28 speakers (14 male/14 female) and the background data of 10 unique background types. Each noise segment is used to generate four files with 0, 5, 10, and 15dB SNR. The published files are sampled at 48kHz and normalized so that the clean speech files have a maximum absolute amplitude of 0.5. We resample them to 16kHz. The complete dataset comprises $11@572$ files.

\mypara{Training}
Network weights and biases are initialized using the Xavier initialization and to zero, respectively. The adaptive normalization parameters are initialized at $\alpha = 1 $ and $\beta = 0$. The feature loss is computed using the first $M=6$ layers. We use the Adam optimizer with a learning rate of $10^{-4}$. We train for 320 epochs (\SI{80}{\hour}) on a Titan X GPU. In each epoch, we present the entire dataset in randomized order ($1$ file per iteration) and files are presented in their entirety.

\section{Experimental Setup}
\label{sec:experiments}

\subsection{Baselines}
\label{ssec:baselines}

As baselines, we use a Wiener filtering pipeline with a priori noise SNR estimation (as implemented in~\cite{HuLoizou2006}), and two recent state-of-the-art methods that use deep networks to perform end-to-end denoising directly on the raw waveform: the Speech Enhancement Generative Adversarial Network (SEGAN)~\cite{Pascual2017} and a WaveNet-based network~\cite{Rethage2017}. This last one is designed around minor modifications to the architecture in~\cite{Oord2016}. It uses stacked context aggregation modules with gated activation units, skip connections, and a conditioning mechanism. The modifications include training with a regression loss ($L^1$ on the raw waveform) rather than a classification loss. The number of layers is larger than in our network (30), while the receptive field is smaller ($3\cdot 2^{11}$ samples), capturing contextual information on more limited time scales. The network architecture is also distinctly more complex than ours. For both deep learning baselines, we use the code and models published by their respective authors. These models are optimized by their authors on the exact same training dataset, allowing fair comparison.

\subsection{Data}
\label{ssec:data}

\begin{figure}[!tb]
	\centering
	\includegraphics[trim={.2cm .3cm .3cm .38cm},clip]{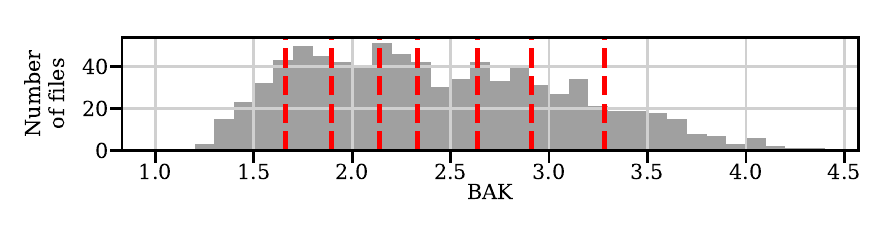}
	\vspace{-3.5mm}
	\caption{Distribution of the test set in terms of composite background score. The test set was partitioned into 8 tranches, demarcated by red dashed lines.}
	\label{fig:distribution}
	\vspace{-4mm}
\end{figure}

\begin{table}[b]
	\centering
	\setlength{\tabcolsep}{3mm}
	\ra{1}
	\vspace{-5mm}
	\caption{Performance for different approaches according to objective quality measures. (Higher is better.)}
	\vspace{-3mm}
	\begin{tabular}{l@{~\quad}c@{~\quad}c@{~\quad}c@{~\quad}c}
		\toprule
		& SNR & SIG & BAK & OVL \\ \midrule
		Noisy  & 8.45 & 3.34 & 2.44 & 2.63 \\
		Wiener  & 12.28 & 3.23 & 2.68 & 2.67 \\
		SEGAN  & 14.82 & 3.21 & 2.76 & 2.56 \\
		WaveNet  & 18.18 & 2.87 & 3.08 & 2.43 \\
		Ours & \bf 19.00 & \bf 3.86 & \bf 3.33 & \bf 3.22 \\ \bottomrule
	\end{tabular}
	\label{tab:results}
\end{table}

All our testing is done in mismatched conditions. The data source is the same as in Section~\ref{ssec:setrain}. The speech is obtained from 2 speakers (1 male/1 female). The background data is obtained from 5 distinct background types. Neither the speakers nor the backgrounds used at test time were seen during training. Each background segment is used to generate four files with 2.5, 7.5, 12.5, and 17.5~dB SNR. The complete test set comprises 824 files. Our denoising pipeline needs about \SI{12}{\milli\second} to process every \SI{1}{\second} of audio in our configuration. The denoised files for our pipeline and the baselines are available as supplementary material at \url{http://ieeexplore.ieee.org}.

\subsection{Quantitative measures}
\label{ssec:metrics}

\paragraph{Objective quality metrics} To evaluate each system, we compare its output to the ground-truth speech signal (i.e., the clean speech alone). The common metrics to measure speech quality given ground-truth are compared in \cite{Loizou2013}. We use here the composite scores from \cite{HuLoizou2006} that were found to be best correlated with human listener ratings. These consist of the overall (OVL), the signal (SIG), and the background (BAK) scores, each on a scale from 1.0 to 5.0, and corresponding respectively to the measure of overall signal quality, the measure of quality when considering speech signal degradation alone, and the measure of quality when considering background signal intrusiveness alone~\cite{ITUT2003}. We also report the SNR \cite{Quackenbush1988}, as a raw measure of the relative energies of the residual background and the speech in a given signal, quantified in decibel~(dB). We use the implementations in~\cite{Loizou2013}. For all metrics, higher scores denote better performance.

The test dataset is divided into 4 mixing SNR subgroups (see Section~\ref{ssec:data}). We argue that the dataset should be rather considered as a continuous distribution of degradation, since SNR correlates poorly with human perception of the degradation level~\cite{Loizou2013}. The continuum of degradation levels is better represented in the distribution of the background intrusiveness BAK score. (The SIG score is less informative since the undistorted speech signal is added.) To evaluate performance as a function of input degradation magnitude, we partition the test set into 8 tranches of equal size, corresponding to the 8 octiles of the BAK score distribution as shown in Figure~\ref{fig:distribution}, with tranches representing a different denoising difficulty.

\begin{figure}[tb]
	\centering
	\begin{tabular}{@{}c@{\,}c@{}}
		\includegraphics[trim={.2cm .8cm .3cm .38cm},clip]{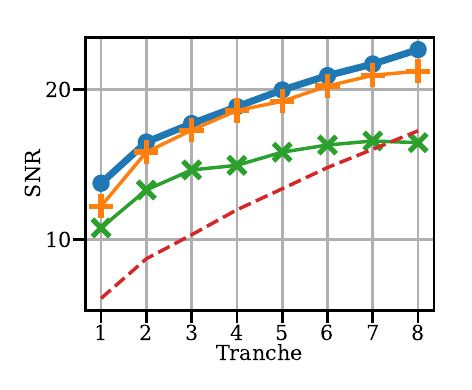}~ &
		\includegraphics[trim={.2cm .8cm .3cm .38cm},clip]{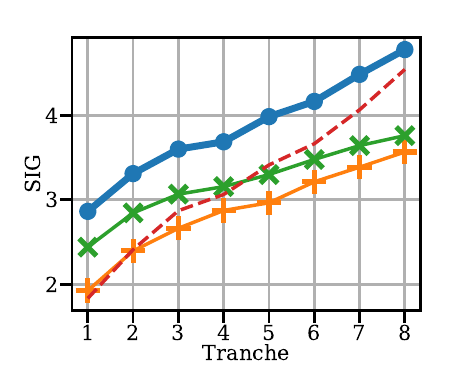} \\[-.9mm]
		\includegraphics[trim={.2cm .3cm .3cm .38cm},clip]{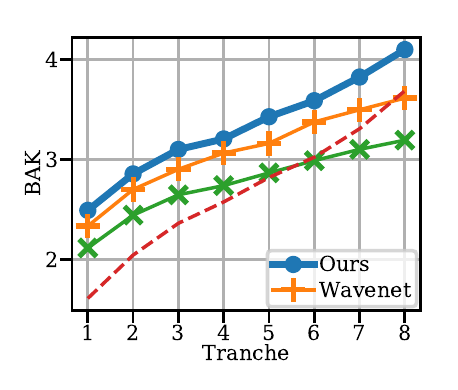} &
		\includegraphics[trim={.2cm .3cm .3cm .38cm},clip]{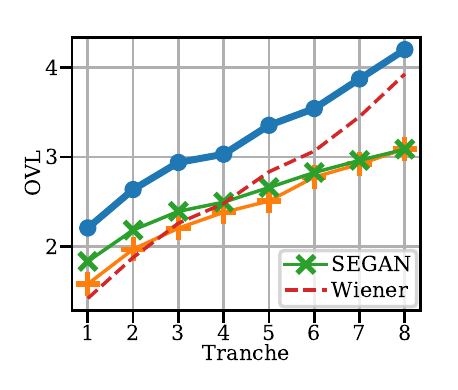}
	\end{tabular}
	\vspace{-2.5mm}
	\caption{Performance of different denoising approaches according to 4 objective quality measures (SNR, SIG, BAK, and OVL), plotted for each tranche in the test set. For all measures, higher is better.}
	\label{fig:results}
	\vspace{-5mm}
\end{figure}

\paragraph{Results} Table~\ref{tab:results} reports these metrics for our approach and the baselines, evaluated over the test set. Our method outperforms all the baselines according to all measures by a comfortable margin. The plots in Figure~\ref{fig:results} further show that our network yields the best quality for all levels of background intrusiveness separated in tranches, with a particularly significant margin according to perceptually-motivated composite measures. Table~\ref{tab:resultsLoss} shows the benefit of using a feature loss compared to training the same denoising network, by the same procedure on the same data, using an $L^1$ or an $L^2$ loss. Training with a feature loss outperforms networks trained with other losses. In particular, while an $L^1$ loss achieves a similar SNR score as our feature loss, the feature loss shows definite improvement for the BAK and OVL metrics. It also scores well for the SIG metric, especially in the noisier tranches, demonstrating the ability to capture meaningful features when important cues are hidden in the noise.

\subsection{Perceptual Experiments}

\paragraph{Experimental design} Objective metrics are known to only partially correlate with human audio quality ratings~\cite{Loizou2013}. Hence, we also conduct carefully designed perceptual experiments with human listeners. The procedure is based on A/B tests deployed at scale on the Amazon Mechanical Turk platform. The A/B tests are grouped into Human Intelligence Tasks (HITs). Each HIT consists of 100 ``ours vs baseline'' pairwise comparisons. Each comparison presents two audio clips that can be played in any order by the worker, any number of times. One of the clips is the output of our approach and one is the output of one of the baselines, for the same input from the test set. The files are presented in random order (both within each pair and among pairs), so the worker is given no information as to the provenance of the clips. The worker is asked to select, within each pair, the clip with the cleaner speech. Each HIT includes 10 additional `sentry' comparisons in which the right answer is obvious to guard against negligent or inattentive workers. These sentry pairs are mixed into the HIT in random order. If a worker gives an incorrect answer to two or more sentry pairs, the entire HIT is discarded. Each HIT then contains a total of 110 pairwise comparisons. A worker is given 1 hour to complete a HIT. Each HIT is completed by 10 distinct workers.

\begin{table}[tb]
	\centering
	\caption{Training the same network with different loss functions. For all metrics, higher is better.}
	\setlength{\tabcolsep}{3mm}
	\ra{1}
	\vspace{-3mm}
	\begin{tabular}{ccccc}
		\toprule
		& SNR & SIG & BAK & OVL  \\ \midrule
		Noisy & 8.45 & 3.34 & 2.44 & 2.63 \\
		L2 & 18.46 & 3.70 & 3.21 & 3.07 \\
		L1  & 18.98 & 3.75 & 3.27 & 3.11 \\
		Feature loss  & \bf 19.00 & \bf 3.86 & \bf 3.33 & \bf 3.22 \\
		\bottomrule
	\end{tabular}
	\vspace{-5mm}
	\label{tab:resultsLoss}
\end{table}

\begin{table}[b]
	\centering
	\setlength{\tabcolsep}{1.5mm}
	\ra{1}
	\vspace{-5mm}
	\caption{Results of perceptual experiments. Each cell lists the fraction of blind randomized pairwise comparisons in which the listener rated the output of our approach as cleaner than the output of a baseline. Each row lists results for a specific baseline. Each column list results for a tranche of the testing set. (Chance is at 50\%, higher is better.)}
	\vspace{-3mm}
	\begin{tabular}{@{}lcccc@{}}
		\toprule
		Tranche: & 1 (Hard) & 3 (Medium) & 5 (Easy) & 7 (Very easy)\\
		\midrule
		Ours $>$ Wiener & 96.1\% & 89.4\% & 81.7\% & 90.2\% \\
		Ours $>$ SEGAN & 83.5\% & 70.5\% & 64.1\%& 61.4\% \\
		Ours $>$ WaveNet & 83.9\% & 67.0\% & 61.4\% & 55.8\%\\
		\bottomrule
	\end{tabular}
	\label{tab:resultsAB}
\end{table}

\paragraph{Results} The results are summarized in Table~\ref{tab:resultsAB}. This table presents the fraction of blind pairwise A/B comparisons in which the listener rated a clip denoised by our network as cleaner than the clip denoised by a baseline. The preference rates are presented versus each baseline across 4 tranches. The most notable results are for the hardest tranche, where the output of our approach was rated cleaner than the output of recent state-of-the-art deep networks in more than 83\% of the comparisons. All results are statistically significant with $p < 10^{-3}$. This demonstrates that our algorithm is more robust in this regime, in which degradation from the background signal is much more noticeable, and for which denoising is particularly useful. For easier tranches, with lower levels of degradation in the input, both our method and the baselines generally perform satisfactorily and listeners can experience more difficulty distinguishing between the different processed files, but the preference rate for our approach remains well above chance (50\%), at statistically significant levels, for all baselines across all tranches.

\section{Conclusion}
\label{sec:conclusion}

We presented an end-to-end speech denoising pipeline that uses a fully-convolutional network, using a deep feature loss network pretrained on several relevant audio classification tasks for training. This approach allows the denoising system to capture speech structure at various scales and achieve better denoising performance without added complexity in the system itself or expert knowledge in the loss design. Experiments demonstrate that our approach significantly outperforms recent state-of-the-art baselines according to objective speech quality measures as well as large-scale perceptual experiments with human listeners. In particular, the presented approach is shown to perform much better in the noisiest conditions where speech denoising is most challenging. Our paper validates the combined use of convolutional context aggregation networks and feature losses to achieve state-of-the-art performance.

\clearpage

\bibliographystyle{IEEEtran}
\bibliography{paper}

\clearpage

\appendix

This appendix provides additional details on the denoising and feature loss network architectures presented in Section~\ref{sec:overview}.

\subsection{Denoising Network}

\mypara{Layer structure} We denote the $16$ (consecutive) network layers by $\Lambda^0,\ldots,\Lambda^{15}$. $\Lambda^0$ and $\Lambda^{15}$ are \mbox{1-dimensional} tensors of dimensionality $N\timess 1$ and correspond to the degraded input signal and the enhanced output signal, respectively. The number of samples $N$ is not given in advance. Each intermediate layer $\Lambda^k \in \{\Lambda^1,\ldots,\Lambda^{15}\}$ is a \mbox{2-dimensional} tensor of dimensionality $N \timess W$, where $W$ is the width of (i.e., the number of feature maps in) each layer. For $k=1,\ldots,14$, the content of each intermediate layer $\Lambda^k$ is computed from the previous layer $\Lambda^{k-1}$ via the operation
\begin{equation}
\Lambda^k_i = \Psi\left( \Gamma^k \left(\sum_j \Lambda_j^{k-1} \ast_{r_k} \KK_{i,j}^k \right)\right),
\label{eq:op_denoising}
\end{equation}
where $\Lambda^k_i$ is the \mbox{$i$-th} feature map of layer $\Lambda^k$, $\Lambda^{k-1}_j$ is the \mbox{$j$-th} feature map of layer $\Lambda^{k-1}$, $\KK_{i,j}^k$ is a learned $3 \timess 1$ convolutional kernel, $\Gamma^k$ is the adaptive normalization operator and $\Psi$ is a pointwise nonlinearity. Because of the presence of adaptive normalization, no bias term is used for these layers. The operator $\ast_{r}$ is a dilated convolution~\cite{YuKoltun2016}, i.e.,
\begin{equation}
\left(\Lambda_j \ast_{r} \KK_{i,j}\right) [n]= \sum_{m = -1}^{+1} \KK_{i,j}[m] \Lambda_j[n-r m].
\end{equation}

The dilation factor for the \mbox{$k$-th} layer is set at $r_k = 2^{k-1}$ for $k\in \{1,\ldots 13\}$. Between layer $\Lambda^{13}$ and $\Lambda^{14}$, we do not use dilation (i.e., $r_{14} = 1$). For the output layer $\Lambda^{15}$, we use a linear transformation ($1\timess 1$ convolution with no nonlinearity) in order to synthesize the sample of the output signal so that
\begin{equation}
\Lambda^{15} = \sum_j \Lambda_j^{14} \timess \KK_{j}^{14} + b,
\end{equation}
where $b$ is a learned bias term. The receptive field of the network is $2^{14}+1=16385$ samples.

\mypara{Nonlinear units} For the pointwise nonlinearity $\Psi$, we use the leaky rectified linear unit (LReLU) \cite{Maas2013}:
\begin{equation}
\Psi(x) = \max(\delta x, x) \text{~with~} \delta = 0.2. \label{eq:lrelu}
\end{equation}

\mypara{Adaptive normalization} $\Gamma^k$ corresponds to the adaptive normalization operation described in Section~\ref{ssec:denoising}. For $k\in \{1,\ldots 13\}$, the operator adaptively combines batch normalization and identity mapping as
\begin{equation}
\Gamma^k(x) = \alpha_k x + \beta_k BN(x),
\label{eq:bn}
\end{equation}
where $\alpha_k, \beta_k \in \mathbb{R}$ are learned scalar weights and $BN$ is the batch normalization operator \cite{IoffeSzegedy2015}.

\mypara{Zero padding} Our algorithm uses zero-padding at each layer so that the ``effective'' length of each layer tensor is constant and identical to $N$.

\mypara{Training loss} The network is trained through backpropagation using our deep feature loss as described in Section~\ref{ssec:loss} (see in particular Equation~\ref{eq:perceptualloss}). The feature loss classification network is further detailed in the next section.

\subsection{Feature Loss Network}
\label{ssec:lossdetails}

\mypara{Feature layer structure}
As mentioned in Section~\ref{ssec:loss}, the network is inspired by the VGG architecture from computer vision. We denote its 15 (consecutive) layers by $\Phi^0,\ldots,\Phi^{14}$. The first layer $\Phi^{0}$ is a \mbox{1-dimensional} tensor of dimensionality $N\timess 1$ and corresponds to the input signal. The number of samples $N$ is not given in advance. Each intermediate layer $\Phi^m\in\{\Phi^1,\ldots,\Phi^{14}\}$ is a \mbox{2-dimensional} tensor of dimensionality $\frac{N}{2^m}\timess W_m$, where $W_m$ is the width of each layer, set to $W_m = 32 \timess 2^{\left \lfloor{\frac{m-1}{5}}\right \rfloor}$ (i.e., the number of features is doubled every 5 layers). The content of each intermediate layer $\Phi^m$ is computed from the previous layer $\Phi^{m-1}$ through the following operation:
\begin{equation}
\tilde{\Phi}^m_i = \Psi\left( BN \left( \sum_j \Phi_{j}^{m-1} \ast \LL_{i,j}^m \right) \right),
\end{equation}
where $\tilde\Phi^m_i$ is the \mbox{$i$-th} feature map of layer $\Phi^m$ prior to the decimation operation, $\Phi^{m-1}_j$ is the \mbox{$j$-th} feature map of layer $\Phi^{m-1}$, $\LL_{i,j}^m$ is a learned $3 \timess 1$ convolutional kernel, $BN$ is the batch normalization operator, and $\Psi$ is the same pointwise linearity as in Equation~\ref{eq:lrelu}. Because of the presence of batch normalization, no bias term is used for these layers. This is followed by the decimation operation
\begin{equation}
\Phi^m_i[n] = \tilde{\Phi}^m_i[2n],
\end{equation}
following which the length of the subsequent layer is half the length of the preceding one. The receptive field of the network is $2^{15}-1=32767$ samples. The network is zero-padded as necessary for each layer so that $\tilde{\Phi}^m$ and $\Phi^{m-1}$ have the same ``effective'' length.

\mypara{Classification layer} To perform the \mbox{$p$-th} classification task of interest, we first average-pool each channel in the last feature layer $\tilde\Phi^{14}$ to yield an output feature vector $\Phi^{15,p}$ of dimensionality $1 \timess W_{14}$. This vector is fed to a linear layer to form a logit vector $\Phi^{15,p}$ of dimensionality $1\timess C_p$ (with $C_p$ the number of classes associated with the \mbox{$p$-th} task) such that
\begin{equation}
\Phi^{16,p}_i = \sum_j \Phi^{15,p}_j \timess \LL^{16,p}_{i,j} + \tilde{b}^p_i,
\end{equation}
where $\tilde\LL^{16,p}_{i,j}$ is a learned scalar weight and $\tilde{b}^p_i$ is a learned bias term. We finally get the output classification vector $\Phi^{17,p}$ of the network through the operation
\begin{equation}
\Phi^{17,p} = \Delta(\Phi^{16,p}),
\end{equation}
where $\Delta$ is the logistic nonlinearity associated with the type of multi-label classification for the \mbox{$p$-th} task (i.e., vector softmax nonlinearity if the task asks for a unique label for each audio file, pointwise sigmoid if the task allows for any number of labels for each audio file). $\Phi^{17,p}$ is of dimension $1\timess C_p$ and its elements are in the range $[0,1]$.

\mypara{Training loss} Training is done through backpropagation using a cross-entropy loss between the vector $\Phi^{17,p}$ associated with the current file (for task $p$) and its corresponding ground truth classification vector (i.e., the vector of dimension $1\timess C_p$ in which the \mbox{$c$-th} element is 1 if the \mbox{$c$-th} classification label is associated with the file, 0 otherwise).

\end{document}